\documentclass[%
 reprint,
 amsmath,amssymb,
 aps,
]{revtex4-2}
\usepackage{float}
\usepackage{graphicx}
\usepackage{dcolumn}
\usepackage{bm}
\usepackage{color}
\usepackage{amsmath}

\newcommand{\vc}{\mathbf}
\usepackage{array}
\newcolumntype{P}[1]{>{\centering\arraybackslash}p{#1}}

\begin{document}

\preprint{APS/123-QED}

\title{Intrinsic bulk viscosity of the one-component plasma}

\author{Jarett LeVan}
\author{Scott D.~Baalrud}
\email{baalrud@umich.edu}
\affiliation{%
Department of Nuclear Engineering and Radiological Sciences, University of Michigan, Ann Arbor, MI 48109, USA
}%

\date{\today}

\begin{abstract}
Intrinsic bulk viscosity of the one-component plasma (OCP) is computed and analyzed using equilibrium molecular dynamics simulations and the Green-Kubo formalism. It is found that bulk viscosity exhibits a maximum at $\Gamma \approx 1$, corresponding to the condition that the average kinetic energy of particles equals the potential energy at the average inter-particle spacing. The weakly coupled and strongly coupled limits are analyzed and used to construct a model that captures the full range of coupling strengths simulated: $\Gamma \approx 10^{-2} - 10^2$. Simulations are also run of the Yukawa one-component plasma (YOCP) in order to understand the impact of electron screening. It is found that electron screening leads to a smaller bulk viscosity due to a reduction in the excess heat capacity of the system. Bulk viscosity is shown to be at least an order of magnitude smaller than shear viscosity in both the OCP and YOCP. The generalized frequency-dependent bulk viscosity coefficient is also analyzed. This is found to exhibit a peak near twice the plasma frequency in strongly coupled conditions, which is associated with the oscillatory decay observed in the bulk viscosity autocorrelation function. The generalized shear and bulk viscosity coefficients are found to have a similar magnitude for $\omega \gtrsim 2\omega_p$ at strongly coupled conditions. 

\end{abstract}

\maketitle

\section{Introduction}
When a fluid is compressed, work is generally done on translational degrees of freedom, creating a temporary imbalance in the energy partition with internal degrees of freedom. Relaxation processes inevitably ensue to restore equilibrium, which result in an irreversible resistance to the compression. An equivalent process exists for expansion, where a fluid does work at the expense of translational kinetic energy leading to an irreversible resistance to the expansion. Hence, fluids dissipate energy when undergoing dilatational motion. To account for this, one must specify a coefficient of bulk viscosity in the Navier-Stokes equations~\cite{landau_lifshitz}. In this work, we study the coefficient of bulk viscosity in the one-component plasma (OCP), a model system often used to study strongly coupled Coulomb systems~\cite{Baus_Hansen_1980}.

Bulk viscosity has historically received considerably less attention than its counterpart shear viscosity, which acts to resist shearing motion. This may, in part, be a result of Stokes himself assuming that the coefficient of bulk viscosity can be well-approximated as zero in most fluids~\cite{Stokes_1845}. It took nearly one-hundred years following Stokes' 1845 paper on viscous fluids for experiments in acoustics to make clear that sound absorption cannot be understood without a proper coefficient of bulk viscosity~\cite{Herzfeld_Rice_1928, Tisza_1942}. Since then, bulk viscosity has also been shown to impact shock wave structure~\cite{Emanuel_1994, Elizarova_2007, Kosuge_2018}, turbulence~\cite{Pan_2017, Chen_2019, Touber_2019}, and instabilities~\cite{Sengupta_2016, Singh_2021}, among other phenomena~\cite{Nazari_2018, Emanuel_1992, Gonzalez_1993, Cunha_2024}.

Bulk viscosity has historically been split into two separate contributions: ``apparent'' bulk viscosity, which arises from relaxation with molecular degrees of freedom (rotation and vibration), and ``intrinsic'' bulk viscosity, which arises from relaxation of the internal potential energy~\cite{Herzfeld_Rice_1928, Hall_1948, Sharma_review}. Of the two, intrinsic bulk viscosity has received considerably less attention. Though often small, it can be significant in dense fluids~\cite{Gray_1964, Hoheisel_1987}. It is also the only mechanism by which a monatomic fluid can exhibit bulk viscosity, the existence of which has been doubted in the past~\cite{Hak_1995,Sharma_2023}.

In this work, we analyze and expand recently obtained molecular dynamics (MD) data of bulk viscosity in the OCP~\cite{LeVan_2024}. When applied to plasmas, the OCP model is composed of ions, each with charge $q$ and mass $m$, at a collective density $n$ and temperature $T$. Electrons are taken to provide a non-interacting, charge neutralizing background. The ions interact through the Coulomb potential
\begin{equation}
    \phi(r) = \frac{q^2}{4 \pi \epsilon_0 r}.
\end{equation}
A particularly attractive feature of the OCP is that when time
is quantified in dimensionless units of the plasma period $\omega_p^{-1} = (\epsilon_0 m / q^2 n)^{1/2}$ and space in units of the average interparticle spacing $a = (3/4\pi n)^{1/3}$, it is entirely characterized by the Coulomb coupling parameter $\Gamma$, defined as
\begin{equation}
    \Gamma = \frac{q^2}{4 \pi \epsilon_0 a k_B T}.
\end{equation}
Understanding transport processes in the OCP directly informs other more complicated systems, such as dense plasmas~\cite{Bergeson_2019,Stanek_2024}

Transport coefficients in the OCP have been heavily studied for decades~\cite{Hansen_1975,Bernu_1978,Donko_2000,Murillo_2000,Saigo_2002,Donko_2008,Khrapak_2013,Strickler_2016,Arkhipov_2017,Khrapak_2018,Scheiner_2019,Kahlert_2024}, but well-resolved bulk viscosity data was published for the first time in 2024~\cite{LeVan_2024}. This is likely because an early MD study showed the bulk viscosity is much smaller than shear viscosity~\cite{Vieillefosse_Hansen_1975}, so it was deemed unimportant for study. A similar result was also found for the magnetized OCP~\cite{Scheiner_2020}. However, bulk viscosity is still not fully understood, and analysis of the OCP provides valuable insight into the underlying mechanism of intrinsic bulk viscosity. Indeed, we find that existing formulas in literature for intrinsic bulk viscosity cannot fully capture the strongly coupled limit of the OCP. However, by analyzing the weakly coupled and strongly coupled limits, we are able to develop a new expression that captures the full range of coupling strengths. Additionally, we note that many fluid simulation algorithms use an artificial bulk viscosity value to ensure numerical stability~\cite{Cook_2005, Kawai_2010, Campos_2018}. It is important, then, to quantify the physical value to confirm that the artificial values used in fluid codes do not miss the associated physical viscous dissipation.   

We find that bulk viscosity in the OCP peaks at $\Gamma \approx 1$, decreasing in the weakly coupled ($\Gamma \to 0$) and strongly coupled ($\Gamma \to \infty$) limits. It is shown that intrinsic bulk viscosity is associated with the relaxation of the radial distribution function, which characterizes the internal potential energy, and that the characteristic relaxation time approaches a constant of the plasma period in the weakly coupled and strongly coupled limits. Specifically, the relaxation time takes a value of $\tau \approx 0.38 \omega_p^{-1}$ in the weakly coupled limit and $\tau = 0.88 \omega_p^{-1}$ in the strongly coupled limit. Across all coupling strengths, the bulk viscosity ($\eta_v$) is found to be at least an order of magnitude smaller than the shear viscosity ($\eta$). 
This implies that macroscopic dynamical behaviors that depend on the longitudinal viscosity $b= \frac{4}{3} \eta + \eta_v$, such as sound wave dissipation~\cite{Hansen_McDonald}, are not significantly influenced by bulk viscosity in the OCP. 

Molecular dynamics simulations of the Yukawa one-component plasma (YOCP), a model system where the ion-ion interaction is screened by electrons, were also run. Electron screening was found to lead to a smaller bulk viscosity due to a weaker interaction between ions. This reduces the excess heat capacity in the YOCP system~\cite{Khrapak_2015}, and therefore the bulk viscosity. 
The bulk to shear ratio is found to be small in the YOCP as well. 

In analyzing the MD data, it is found that the relaxation time of the autocorrelation function associated with bulk viscosity is much faster than that of shear viscosity. For this reason, the frequency-dependent bulk and shear viscosities were evaluated to determine if, at sufficiently fast timescales, bulk viscosity may be the dominant form of energy dissipation. This was found to not be the case, as bulk viscosity becomes larger than shear viscosity only on timescales at which both transport coefficients are negligible. 

This paper is organized as follows: Section \ref{sec-two} reviews some of the physical intuition behind bulk viscosity and existing expressions for intrinsic bulk viscosity. In Section \ref{sec-three}, we analyze MD data for the OCP bulk viscosity, provide an accurate model, and show MD results of the YOCP. In Section \ref{sec-freq}, the frequency dependence of bulk viscosity is presented and analyzed. Concluding comments are provided in Section \ref{sec-conclusion}.   

\section{Intrinsic Bulk Viscosity \label{sec-two}}

\subsection{Definitions}

The stress tensor ($\bm{\Pi}$) for a compressible Newtonian fluid is~\cite{Hansen_McDonald}
\begin{equation}
    \bm{\Pi} = -p_t \vc{I} + \eta \biggl[ \nabla \vc{V} + (\nabla \vc{V})^T - \frac{2}{3} (\nabla \cdot \vc{V}) \vc{I} \biggr] + \eta_v (\nabla \cdot \vc{V}) \vc{I},
    \label{eq-stress}
\end{equation}
where $p_t$ is the thermodynamic pressure given by the equation of state, $\vc{V}$ is the fluid velocity, $\eta$ is the coefficient of shear viscosity, and $\eta_v$ is the coefficient of bulk viscosity.   
The stress tensor appears in the hydrodynamic equation for momentum conservation, 
\begin{equation}
    \rho \frac{d\vc{V}}{dt} = \nabla \cdot \bm{\Pi}
\end{equation}
where $d/dt = \partial/\partial t + \vc{V} \cdot \nabla$ is the convective derivative and $\rho$ is the mass density, as well as energy conservation 
\begin{equation}
    \frac{du}{dt} = \lambda \nabla^2 T + \bm{\Pi} : \nabla \vc{V} ,
\end{equation}
where $u$ is the energy density and $\lambda$ is the thermal conductivity. The two previous equations, combined with the mass conservation equation  
\begin{equation}
    \frac{d\rho}{dt} = -\nabla \cdot (\rho \vc{V}),
\end{equation}
and an equation of state, form the complete set of Navier-Stokes equations~\cite{Hansen_McDonald}. It is through these equations that bulk viscosity affects the flow of a fluid. 

To elucidate the physical meaning of intrinsic bulk viscosity, consider the instantaneous (mechanical) pressure of a fluid, $p_m$, defined as $p_m = -\frac{1}{3}\textrm{Tr}\lbrace \bm{\Pi} \rbrace$. Plugging into Eq.~(\ref{eq-stress}) yields
\begin{equation}
    p_m - p_t = -\eta_v \nabla \cdot \vc{V}.
    \label{eq-dp}
\end{equation}
This expression shows that bulk viscosity determines the extent to which a fluid's  mechanical pressure is away from its equilibrium value during expansion or compression. 
It can alternatively be written in terms of the radial distribution function, $g(r)$. 
To see this, first consider the relationship between pressure, temperature, interaction potential and $g(r)$ from classical statistical mechanics~\cite{Hansen_McDonald} 
\begin{subequations}
\begin{align}
    p_m &= nk_B T - \frac{2}{3} \pi n^2 \int_0^\infty \frac{d\phi}{dr} r^3 g(r) dr, \label{eq-pm}\\
    p_t &= nk_B T_{\mathrm{eq}} - \frac{2}{3} \pi n^2 \int_0^\infty \frac{d\phi}{dr} r^3 g_{\mathrm{eq}}(r) dr   
    \label{eq-pt}
\end{align}
\end{subequations}
where $n$ is the number density and ``eq'' denotes the equilibrium value of the particular quantity. Next, the conservation of energy 
\begin{equation}
    \frac{3}{2} n k_B (T - T_{\mathrm{eq}}) + 2 \pi n^2 \int_0^\infty \phi(r) r^2 \left[g(r) - g_{\mathrm{eq}}(r)\right] dr = 0
\end{equation}
can be used to relate the temperature difference between the current and equilibrium state to the potential energy difference between the two states.
Plugging the energy conservation relation into Eq.~(\ref{eq-dp}) provides 
\begin{equation}
   \eta_v \nabla \cdot \vc{V} = \frac{2}{3}\pi n^2 \int_0^\infty \left[ r \frac{d \phi(r)}{dr} + 2 \phi(r) \right] \delta g(r) r^2 dr ,
    \label{eq-dp}
\end{equation}
where $\delta g(r) \equiv g(r) - g_{\mathrm{eq}}(r)$. 
From this expression, it is seen that intrinsic bulk viscosity has its origins in the system's radial distribution function $g(r)$ being away from its equilibrium form. The origin of this non-equilibrium is simple. When a fluid is compressed, it takes a finite amount of time for its constituent particles to move to their minimum potential energy state. During that time, $g(r) \neq g_{\mathrm{eq}}(r)$. Of course, the longer that relaxation time, the larger the deviation from equilibrium and hence a larger bulk viscosity. 

\begin{figure}
    \centering
    \includegraphics[]{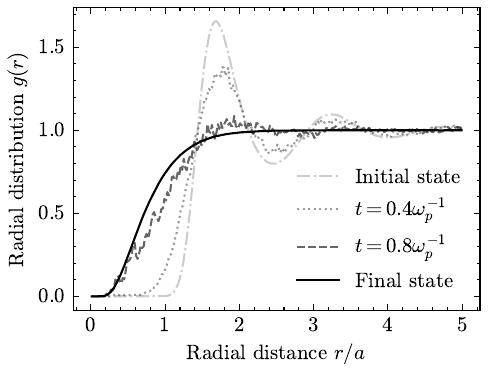}
    \caption{Time evolution of the radial distribution function $g(r)$ following a sudden increase of the system's translational energy.}
    \label{fig-gr}
\end{figure}

An example of the $g(r)$ relaxing in response to a perturbation is shown in Fig.~\ref{fig-gr}. For this illustration, a rather large perturbation, consisting of a sudden change from $\Gamma = 50$ to $\Gamma = 1$ by raising the temperature, was used to make the $g(r)$ evolution plainly visible. From this, it is clearly seen that the $g(r)$ approaches its equilibrium form over the course of about one plasma period. This rate of $g(r)$ relaxation has been observed previously in disorder-induced heating~\cite{Killian_2007,Acciarri_2022,LeVan_DIH}.  
Since Eq.~(\ref{eq-dp}) connects the intrinsic bulk viscosity with the evolution of $g(r)$, we expect this to define the relaxation timescale of pressure perturbations that lead to the bulk viscosity. 

\subsection{First-Principles Methods}

Bulk viscosity can be computed from particle data generated by MD simulations using a variety of methods. The most popular method is to run equilibrium MD simulations and extract bulk viscosity from the Green-Kubo relation~\cite{Green_1952, Kubo_1957, Hansen_McDonald}
\begin{equation}
    \eta_v = \frac{\mathcal{V}}{k_B T} \int_0^\infty \langle \delta p(t) \delta p(0) \rangle dt
    \label{eq-gk}
\end{equation}
where $\delta p (t) = p_m(t) - p_t$, $\mathcal{V}$ is the volume, and the angle brackets $\left< \ldots \right>$ denote an ensemble average. 
Here, the mechanical pressure is computed from the stress tensor obtained from individual particle trajectories~\cite{Hansen_McDonald}
\begin{equation}
    \Pi_{ij} = -\frac{1}{\mathcal{V}} \sum_{\alpha=0}^{N_\alpha} \left[ m_\alpha v_{\alpha_i}v_{\alpha_j} + \sum_{\beta>\alpha}^{N_\alpha} (r_{\alpha \beta_i} F_{\alpha \beta_j})\right],
    \label{eq:md-stress}
\end{equation}
where $r_{\alpha \beta}$ is the distance between particles $\alpha$ and $\beta$, and $F_{\alpha \beta}$ is the force between them. 
Specifically, the mechanical pressure is computed from $p_m(t) = -\frac{1}{3}\textrm{Tr}\lbrace \bm{\Pi}(t) \rbrace$ and the thermodynamic pressure $p_t$ from averaging $p_m(t)$ over a thermodynamically long timescale. 
This approach is powerful because Eq.~(\ref{eq-gk}) can be computed directly from equilibrium MD simulations. The bulk viscosity of a wide variety of gases and liquids \cite{Guo_2001,Meier_2004,Jaeger_2018,Sharma_2022}, as well as plasmas~\cite{Vieillefosse_Hansen_1975,Scheiner_2020,LeVan_2024} have been calculated using it. However, it should be noted that evaluating the Green-Kubo relation is often computationally expensive since long MD simulations are required for the integral to converge. 

Methods of computing bulk viscosity from non-equilibrium MD have also been developed. In Hoover \textit{et al}'s method~\cite{Hoover_1980}, the simulation box is cyclically expanded and compressed as
\begin{equation}
    L(t) = L_0 + \Delta L \sin \omega t,
\end{equation}
where $L(t)$ is the length of any one side of the box, $L_0$ the average length, $\Delta L$ the maximum change in $L$, and $\omega$ the frequency at which the box oscillates. The bulk viscosity is computed by analyzing the energy increase per cycle ($E_c$) as
\begin{equation}
    \eta_v = \frac{E_c L_0^2}{9 \pi \omega \mathcal{V} \Delta L^2}.
\end{equation}
In Heyes' method~\cite{Heyes_1984}, the simulation box in an MD simulation is instantaneously and isotropically changed from a volume $\mathcal{V}$ to $\mathcal{V} + \Delta \mathcal{V}$, and the bulk viscosity then evaluated as
\begin{equation}
    \eta_v = -\frac{\mathcal{V}}{\Delta \mathcal{V}} \int_0^\infty  \delta p(t) dt ,
\end{equation}
where the mechanical pressure is again given by $p_m(t) = -\frac{1}{3}\textrm{Tr}\lbrace \bm{\Pi}(t) \rbrace$, but the thermodynamic pressure is given by $p_t = \lim_{t\to\infty} p_m(t)$, i.e., the pressure of the system after it has fully relaxed.
In Sharma's method, the simulation box in an MD simulation is expanded isotropically at a constant rate $\nabla \cdot \vc{V}$, and bulk viscosity evaluated from the pressure difference as \cite{Sharma_2019}
\begin{equation}
    \eta_v = -\delta p / \nabla \cdot \vc{V},
\end{equation}
where $\delta p(t)$ is computed using $\delta p(t) = 2(E_{\mathrm{trans}} - \frac{6}{f}E)/(3 \mathcal{V})$ 
where $f$ is the number of degrees of freedom, $E_{\mathrm{trans}}$ is the translational energy, and $E$ is the total energy. This approach only considers the apparent bulk viscosity mechanism.

The non-equilibrium methods offer unique insight into the underlying microscopic physics at work. However, they also can be computationally expensive, often more so than the Green-Kubo method, and a single simulation only gives the coefficient of bulk viscosity for one particular set of conditions. For this reason, there is much interest in a formula that can be evaluated from theory alone. 

\subsection{Approximate Methods}

Okumura \textit{et al}~\cite{Okumura_2002} developed an expression for bulk viscosity by assuming that the time-evolution of $g(r)$ during compression can be written as
\begin{equation}
    \left( \frac{\partial g(r, t)}{\partial t} \right)_r = - \frac{\dot{L}}{L} r \left( \frac{\partial g(r, t)}{\partial r}\right)_t - \frac{1}{\tau} \left[ g(r, t) - g_0(r, \mathcal{V}(t))\right]
    \label{eq-relax}
\end{equation}
where $L$ is the length of one side of the box, $g_0(r, \mathcal{V}$) is the equilibrium radial distribution function, and $\tau$ is the relaxation time of $g(r)$. The first term on the right hand side is due to uniform compression, and the second term represents the effect of bulk viscosity.
They conducted a perturbation expansion of the non-equilibrium $g(r, t)$ about its equilibrium value $g_0(r, \mathcal{V}(t))$ to obtain
\begin{align}
\label{eq-okumura}
    \eta_v &= -\frac{n^2}{18} \tau \int_0^\infty \left[ r \frac{d \phi(r)}{dr} + 2 \phi(r)\right] \times \\ \nonumber  &\left[ r \left( \frac{\partial g_0 (r, \mathcal{V})}{\partial r}\right)_{\mathcal{V}, S} + 3 \mathcal{V} \left( \frac{\partial g_0(r, \mathcal{V})}{\partial \mathcal{V}} \right)_{r, S} \right] 4 \pi r^2 dr
\end{align}
where $\phi(r)$ is the pair potential between particles and $S$ is the entropy.

This formula has been applied to Lennard-Jones fluids and shown to provide accurate results in comparison to MD simulations~\cite{Okumura_2002, Zaheri_2007}. However, its application utilizes MD simulations to determine $\tau$ and $(\partial g_0(r, \mathcal{V})/\partial \mathcal{V} )_{r, S}$, since it can be difficult to determine how $g(r)$ changes isentropically with volume. 

In the next section, it is shown that Okumura's formula predicts systems that interact through a repulsive potential have a negative bulk viscosity. However, if the sign is changed, the formula gives a good approximation for the magnitude of bulk viscosity in the OCP, with some error at strong coupling. In order to better understand intrinsic bulk viscosity, a new formula is presented that can match the MD data across all coupling strengths.

\section{Bulk viscosity of the one-component plasma 
\label{sec-three}}

\subsection{Molecular Dynamics Results \label{sec-md}}

Recent work used the Green-Kubo relation from Eq.~(\ref{eq-gk}) to calculate the bulk viscosity of the OCP for $\Gamma$ values of $1 - 100$ \cite{LeVan_2024}. Here, we extend this dataset down to $\Gamma = 0.01$ to understand the limit of weak coupling, and obtain more accurate data for $\Gamma = 1 - 100$. Molecular dynamics simulations were run using the open-source code LAMMPS \cite{LAMMPS}. The timestep and number of particles for each simulation is given in Tbl.~\ref{tab-md-param}. The particle-particle-particle-mesh (P$^3$M) algorithm was used, where short-range interactions ($r < 5a$) are computed exactly from the Coulomb force, while long-range interactions ($r > 5a$) are computed by interpolating the electric fields to a mesh~\cite{Hockney1966ComputerSU}. 

Simulations started by placing a fixed number of particles in a box with periodic boundary conditions, then advancing the dynamics in a canonical (NVT) ensemble using a Nos\'{e}-Hoover thermostat to establish equilibrium at a desired temperature. The equations of motion were then evolved in the microcanonical (NVE) ensemble and the stress tensor computed from individual particle trajectories using Eq.~(\ref{eq:md-stress}) output every 10 timesteps in order to compute the Green-Kubo relation using Eq.~(\ref{eq-gk}). The quantity $p_m(t) = -\frac{1}{3}\textrm{Tr}\lbrace \bm{\Pi}(t) \rbrace$ was output as a function of time, and $p_t$ was obtained from the average of this over the entire simulation time. 

\begin{table}[]
    \centering
    \begin{tabular}{c|c|c}
        Coupling parameter & Timestep ($\omega_p^{-1}$) & Number of particles \\ \hline
        $\Gamma < 0.1$ & $10^{-4} $ & 50000\\
        $0.1 \leq \Gamma < 1$ & $10^{-3} $ & 10000\\
        $\Gamma \geq 1$ & $10^{-2}$ & 5000
    \end{tabular}
    \caption{MD simulation details.}
    \label{tab-md-param}
\end{table}

\begin{figure}
    \centering
    \includegraphics[]{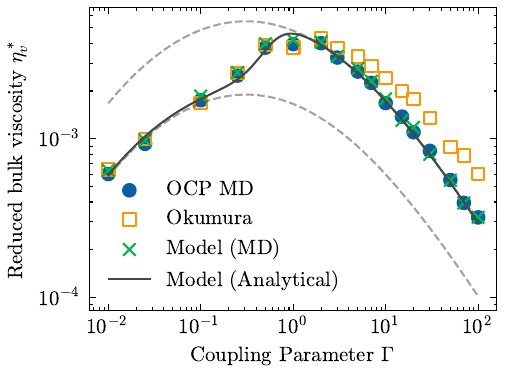}
    \caption{Reduced bulk viscosity of the OCP as a function of $\Gamma$. Data includes MD evaluation of the Green-Kubo relation (blue circles), evaluation of Okumura's formula (Eq.~(\ref{eq-okumura})) using the HNC approximation for $g(r)$ and MD data for $\tau$ (orange squares), evaluation of the model from Eq.~(\ref{eq-model}) using MD data for $\tau$ and $c_v$ (green x's), and evaluation of the model from Eq.~(\ref{eq-model}) using the fit to HNC heat capacity data and the $\sigma$ function to bridge strong and weak coupling limits (black line). Dashed lines show the limiting behavior in weak and strong coupling from Eqs.~(\ref{eq-weak-model}) and (\ref{eq-strong-model}), respectively. }
    \label{fig-model}
\end{figure}

\begin{table}[]
    \centering
    \begin{tabular}{c|c| |c|c}
        $\Gamma$ & $\eta_v^*$ & $\Gamma$ & $\eta_v^*$  \\ \hline
        0.01 &  $6.0 \times 10^{-4}$ & 7 & $2.25 \times 10^{-3}$ \\
        0.025 & $9.3 \times 10^{-4}$ & 10 & $1.68 \times 10^{-3}$ \\
        0.1 & $1.75 \times 10^{-3}$ & 15 & $1.38 \times 10^{-3}$ \\
        0.25 & $2.5 \times 10^{-3}$ & 20 & $1.10 \times 10^{-3}$ \\
        0.5 & $3.75 \times 10^{-3}$ & 30 & $8.40 \times 10^{-4}$ \\
        1 & $3.95 \times 10^{-3}$ & 50 & $5.50 \times 10^{-4}$ \\
        2 & $4.0 \times 10^{-3}$ & 70 & $3.95 \times 10^{-4}$\\
        3 & $3.25 \times 10^{-3}$ & 100 & $3.20 \times 10^{-4}$ \\
        5 & $2.65 \times 10^{-3}$ \\
    \end{tabular}
    \caption{Reduced coefficient of bulk viscosity in the OCP computed using equilibrium molecular dynamics under the Green-Kubo formalism.}
    \label{tab:table}
\end{table}

The full set of data is shown in Fig.~\ref{fig-model} and given in Tbl.~\ref{tab:table}. Values are provided in terms of reduced (dimensionless) bulk viscosity, defined as 
\begin{equation}
    \eta_v^* = \frac{\eta_v}{mna^2 \omega_p}.
\end{equation}
The bulk viscosity is observed to peak around $\Gamma = 1$ and approach zero in the limits of both small and large coupling. Physically, this behavior can be understood as follows. The weakly coupled limit corresponds to an ideal gas, which is known to have zero bulk viscosity \cite{chapman1990mathematical}. This occurs because the excess heat capacity of an ideal gas is zero, implying that every possible arrangement of particles is at the minimal potential energy state ($U = 0$). Therefore, when an ideal gas is compressed, no relaxation processes need occur to obtain equilibrium. In terms of Eq.~(\ref{eq-gk}), $g(r)=1$ in a ideal gas regardless of the density, so $\delta g(r) \rightarrow 0$, and therefore $\eta_v \rightarrow 0$. In the strongly coupled limit, the system's kinetic energy is small compared to its internal potential energy. This leads to a decreasing bulk viscosity because pressure fluctuations in the OCP are proportional to $T^2$ (see Sec.~\ref{sec-model}). It is evident that bulk viscosity is maximized when there is a balance between kinetic energy and internal energy, which corresponds to a $\Gamma$ of unity. 

\begin{figure}
    \centering
    \includegraphics[width=8.5cm]{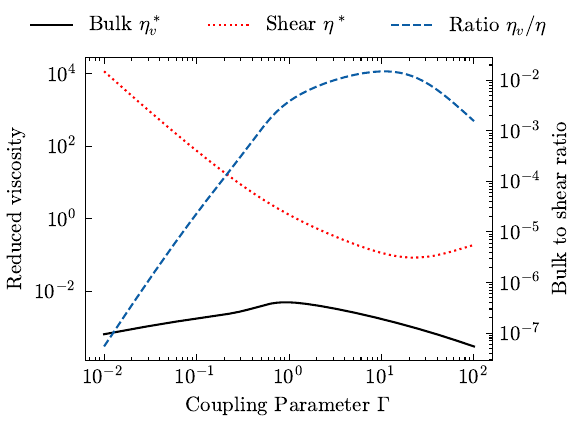}
    \caption{Reduced bulk viscosity and reduced shear viscosity of the OCP computed using Eq.~(\ref{eq-model}) and the fit from Ref.~\cite{Daligault_2014}, respectively. The bulk to shear ratio is also shown.}
    \label{fig-bulktoshear}
\end{figure}

A comparison of bulk viscosity and shear viscosity is shown in Fig.~(\ref{fig-bulktoshear}). Here, the shear viscosity is computed from a fit to MD simulation results by Daligault~\textit{et al}~\cite{Daligault_2014}. It can be seen that bulk viscosity is at least an order of magnitude smaller than shear viscosity across all coupling strengths in the OCP, consistent with previous results~\cite{Vieillefosse_Hansen_1975}. The bulk to shear ratio peaks at $\Gamma = 10$ with a value of $\eta_v / \eta = 1.5 \times 10^{-2}$. 

\subsection{Model \label{sec-model}}

\begin{figure}
    \centering
    \includegraphics[width=8.5cm]{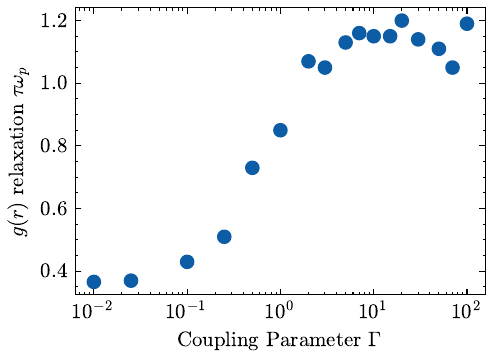}
    \caption{Relaxation time computed using an MD evaluation of Eq.~(\ref{eq-tau-okumura}).}
    \label{fig-taus}
\end{figure}

In an attempt to model the MD data, we evaluated Okumura's formula from Eq.~(\ref{eq-okumura}) using the hypernetted-chain approximation (HNC) of $g(r)$ with a bridge function to extend to high coupling, as described in Refs.~\cite{Ng_1974} and \cite{Iyetomi_1992}. The form of the relaxation of $g(r)$ assumed by Okumura in Eq.~(\ref{eq-relax}) implies that one can compute $\tau$ from~\cite{Zaheri_2007}
\begin{equation}
    \tau = \frac{\int_0^\infty \delta p(t) dt}{\delta p(0)}.
    \label{eq-tau-okumura}
\end{equation}
A plot of $\tau$ as a function of $\Gamma$ computed from MD is shown in Fig.~\ref{fig-taus}. Upon plugging these values into Eq.~(\ref{eq-okumura}), Okumura's formula yields a negative value for bulk viscosity. The formula seems to predict that any system which interacts through a repulsive potential will have a negative bulk viscosity, since the function $(\partial g_0 (r, \mathcal{V})/\partial r)_{\mathcal{V}, S}$ is in general positive and greater than $3 \mathcal{V} \left( \partial g_0(r, \mathcal{V})/\partial \mathcal{V} \right)_{r, S}$, which appears negative in general. However, the magnitude of the bulk viscosity predicted by Okumura's formula matches well with the MD data. This is shown in Fig \ref{fig-model}. Some disagreement is observed in the strong coupling regime, which will be remedied by the alternate formulation proposed below.

An analysis of the $\delta p$ autocorrelation function shows that in the weakly coupled limit there is a characteristic exponential decay, such that
\begin{equation}
    \delta p(t) = \delta p(0) e^{-t/\tau_\textrm{wc}}.
    \label{eq-weak}
\end{equation}
Figure~\ref{fig-taus} shows that the relaxation time $\tau_\textrm{wc}$ converges to a constant of the plasma period at weak coupling of approximately $\tau_\textrm{wc} = 0.38 \omega_p^{-1}$. 
However, in the strong coupling limit the autocorrelation decay is closer to a Gaussian form
\begin{equation}
    \delta p(t) = \delta p(0) e^{-t^2/2\tau_\textrm{sc}^2}, \label{eq-strong}
\end{equation}
where the relaxation time $\tau_\textrm{sc}$ again converges to a constant of the plasma period, but now with a value $\tau_\textrm{sc} = 0.88 \omega_p^{-1}$. Note that this differs from the value observed in Fig.~(\ref{fig-taus}) by a factor of $\sqrt{\pi/2}$ because Okumura's expression assumed an exponential decay of the autocorrelation function as opposed to a Gaussian.

\begin{figure}
    \centering
    \includegraphics[width=8.5cm]{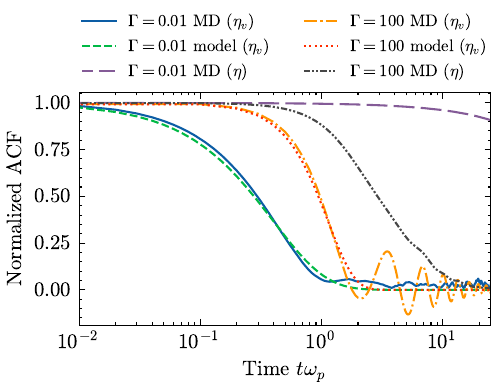}
    \caption{Autocorrelation functions (ACFs) of the bulk viscosity and shear viscosity in the weakly coupled ($\Gamma = 0.01$) and strongly coupled ($\Gamma = 100$) limits. ``Model'' refers to Eq.~(\ref{eq-weak}) for the weakly coupled case and Eq.~(\ref{eq-strong}) for the strongly coupled case. The ACFs are normalized such that their initial value is 1.}
    \label{fig-acfs}
\end{figure}

A comparison of the model and MD data for the decay of the pressure autocorrelation function is shown in Fig. \ref{fig-acfs}. 
The model captures the characteristic decay time in both the weak and strong coupling limits. 
Notably, large oscillations are present in the strong coupling limit that are not captured by the Gaussian form of Eq.~(\ref{eq-strong}). However, as shown below, the approximation is sufficient to accurately capture the bulk viscosity coefficient in comparison to MD. For this reason, we do not attempt to model the oscillations. 

The physical meaning of $\tau$ can be shown by first rewriting Eq.~(\ref{eq-weak}) in differential form as
\begin{equation}
    \frac{d\delta p}{dt} = -\frac{\delta p}{\tau}.
\end{equation}
Then, plugging in the expression for $\delta p$ using Eqs.~(\ref{eq-pm}) and (\ref{eq-pt}), one can obtain
\begin{equation}
    \frac{dg(r, t)}{dt} = -\frac{g(r, t) - g_{\mathrm{eq}}(r)}{\tau}
\end{equation}
by which it is clear that $\tau$ represents the relaxation time of $g(r)$. One can show the same for Eq.~(\ref{eq-strong}), though for the strong coupling case the $g(r)$ relaxation is Gaussian. The differing behaviors in the two limits is not surprising. In the weakly coupled limit, particles experience occasional binary collisions, similar to a neutral gas. On the other hand, in the strongly coupled limit, particles are continuously under the influence of their neighbors, resulting in complex many-body physics. 

One interesting feature of the relaxation of the $\delta p(t)$ autocorrelation function is that $\tau$ approaches a constant fraction of the plasma period in the weakly coupled limit. Typically, the relaxation time of correlations functions associated with transport coefficients follow the associated Coulomb collision rate \cite{Bernu_1978,Daligault_2012,Baalrud_2013}. For example, shear viscosity has a relaxation time that scales as $\tau \omega_p \propto \Gamma^{-3/2}$ in the weakly coupled limit, leading to a relaxation time on the order of $10^{3} \omega_p^{-1}$ when $\Gamma = 0.01$~\cite{Daligault_2014}. 
This is because the shear stress autocorrelation function is determined by velocity relaxation processes in the weakly coupled limit~\cite{Daligault_2014}. 
In contrast, relaxation of $\delta p(t)$, which controls bulk viscosity, is associated with the spatial configuration; $g(r)$. This relaxes at a much faster rate that is a constant multiple of the plasma period;  $\tau_\textrm{wc} = 0.38 \omega_p^{-1}$. This difference in timescales is shown in Fig.~\ref{fig-acfs}. 
Specifically, the decay time of the shear stress autocorrelation function is found to be a few orders of magnitude longer than the pressure autocorrelation function at weak coupling. 
In contrast, the decay times are of the same order at strong coupling because the Coulomb collision time approaches a multiple of the plasma frequency ($\nu \approx 0.2 \omega_p$)~\cite{Baalrud_2013}.

To explain the fast decay time of $g(r)$, we can estimate that in the weakly coupled limit the spatial configuration relaxes on a timescale associated with the time it takes ions to move a correlation length, which is the Debye length $\lambda_D= (\epsilon_0 k_B T / nq^2)^{1/2}$. Taking the characteristic ion speed to be the thermal speed, $v_{\mathrm{th}} = (8k_B T / \pi m)^{1/2}$, the characteristic relaxation time is  
   $ \tau_\textrm{wc} \sim \lambda_D/v_{\mathrm{th}} \approx 0.6 \omega_p^{-1}$.
This estimate produces the correct scaling, though the true relaxation time differs by about 40\%.

Knowledge of the decay time of the autocorrelation function can be used to construct an approximate expression for bulk viscosity. In the weakly coupled limit, putting Eq.~(\ref{eq-weak}) into the Green-Kubo relation from Eq.~(\ref{eq-gk}) provides
\begin{equation}
    \eta_v^\textrm{wc} = \frac{\mathcal{V}}{k_B T} \langle \delta p^2(0) \rangle \tau_\textrm{wc}.
    \label{eq-init}
\end{equation}
Similarly, in the strongly coupled limit, inserting Eq.~(\ref{eq-strong}) into Eq.~(\ref{eq-gk}) provides
\begin{equation}
    \eta_v^\textrm{sc} = \sqrt{\frac{\pi}{2}}\frac{\mathcal{V}}{k_B T} \langle \delta p^2(0) \rangle \tau_\textrm{sc}.
    \label{eq-init-strong}
\end{equation}
A suitable expression for $\langle \delta p^2(0) \rangle$ will therefore provide a model for the bulk viscosity. 

The magnitude of kinetic energy fluctuations in the microcanonical ensemble can be determined exactly \cite{Lebowitz_1967}, so it is desirable to convert $\delta p^2$ to $(\Delta K)^2$, where $K$ is the system's kinetic energy. In a system interacting through a $1/r$ potential, like the OCP, one has the simple relationship between pressure and energy \cite{Baus_Hansen_1980}
\begin{equation}
    p = \frac{2K + U}{3 \mathcal{V}}
\end{equation}
where $U$ is the potential energy. 
Therefore, 
\begin{equation}
    \delta p^2 = \frac{4(\Delta K)^2 + (\Delta U)^2 + 4 \Delta K \Delta U}{9 \mathcal{V}^2} = \frac{(\Delta K)^2}{9 \mathcal{V}^2},
\end{equation}
where the equation on the far-right is obtained by applying conservation of energy, $\Delta K = -\Delta U$. Finally, we make use of the expression for kinetic energy fluctuations found by Lebowitz~\textit{et al}~\cite{Lebowitz_1967}
\begin{equation}
    (\Delta K)^2 = \frac{3}{2}N(k_B T)^2 \left(1 - \frac{3k_B}{2c_v} \right),
    \label{eq-dk}
\end{equation}
where $c_v$ is the heat capacity per particle at constant volume.
With this, Eq.~(\ref{eq-init}) becomes
\begin{equation}
    \eta_{v}^\textrm{wc} = \frac{1}{6}nk_B T \left( 1 - \frac{3k_B}{2c_v} \right) \tau_{\textrm{wc}} ,
    \label{eq-weak-model}
\end{equation}
where $\tau_{\textrm{wc}} = 0.38\omega_p^{-1}$
which is our model for bulk viscosity in the weakly coupled limit. Similarly, Eq~(\ref{eq-init-strong}) becomes
\begin{equation}
    \eta_{v}^\textrm{sc} = \frac{\sqrt{\pi}}{6\sqrt{2}}nk_B T \left( 1 - \frac{3k_B}{2c_v} \right) \tau_\textrm{sc}, 
    \label{eq-strong-model}
\end{equation}
where $\tau_\textrm{sc} = 0.88 \omega_p^{-1}$, which is our model for bulk viscosity in the strongly coupled limit. 
The only unknown in these two equations is $c_v$. To evaluate this, we use the relationship between $c_v$ and $g(r)$ from classical statistical physics~\cite{Hansen_McDonald}
\begin{equation}
    c_v = \frac{3}{2}k_B + 2 \pi n \int_0^\infty \phi(r) \left(\frac{\partial g(r)}{\partial T}\right)_\mathcal{V}r^2 dr
    \label{eq-cv}
\end{equation}
and evaluate $g(r)$ using the HNC approximation with a bridge function, as mentioned previously. 
In order to give an algebraic model, we constructed a fit to the $c_v$ data from HNC that recovers the weakly coupled limit scaling.

The weakly coupled limit can be approximated using the Debye-H\"{u}ckel model  
\begin{equation}
    g(\tilde{r}) = \exp \left( -\frac{\Gamma}{\tilde{r}} e^{-\tilde{r} \sqrt{3 \Gamma}}\right)
\end{equation}
where $\tilde{r} = r/a$ is the dimensionless radial distance. It is desirable to cast Eq.~(\ref{eq-cv}) in dimensionless form in order to compute the temperature derivative in Eq.~(\ref{eq-cv}). Defining $\tilde{c}_v = c_v/k_B$, using 
\begin{equation}
    \left(\frac{\partial g}{\partial T}\right)_\mathcal{V} = -\frac{\Gamma}{T} \frac{\partial g}{\partial \Gamma}
\end{equation}
and the approximation 
\begin{equation}
    \exp \left( - \frac{\Gamma}{\tilde{r}} e^{-\tilde{r}\sqrt{3\Gamma}} - \tilde{r}\sqrt{3\Gamma} \right) \approx \exp \left(-\tilde{r}\sqrt{3\Gamma}\right),
\end{equation}
Eq.~(\ref{eq-cv}) is
\begin{equation}
    \tilde{c}_v^\textrm{wc} = \frac{3}{2} \left(1 + \frac{\Gamma^{3/2}}{2\sqrt{3}} \right) .
\end{equation}
This quantifies the weakly coupled limit of heat capacity at constant volume in the OCP. 

\begin{figure}
    \centering
    \includegraphics[width=8.5cm]{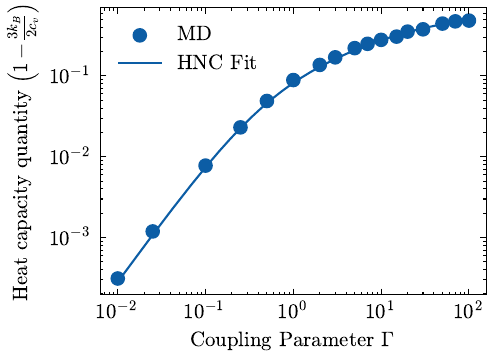}
\caption{The quantity $1 - \frac{3k_B}{2c_v}$ computed from MD using Eq.~(\ref{eq-dk}) (circles) and using the fit to HNC data from Eq.~(\ref{eq-cvfit}) (line).}
    \label{fig-cvs}
\end{figure}

Finally, to obtain an analytic formula for the entire range of coupling strengths, we match the analytic weakly coupled limit with a fit to the numerical data in the strongly coupled limit using a Pad\'{e} approximate
\begin{equation}
    \tilde{c}_v = \frac{3}{2} \left(1 + \frac{\Gamma^{3/2}}{2\sqrt{3}} \frac{a_1 + a_2\Gamma + a_3 \Gamma^2}{a_1 + b_2\Gamma + b_3\Gamma^2} \right)
    \label{eq-cvfit}
\end{equation}
where $a_1 = -1176.6$, $a_2 = 142.97$, $a_3 = -0.40909$, $b_2 = -2477.8$, and $b_3 = 310.57$.
To assess the accuracy of this, heat capacity was computed from MD using Eq.~(\ref{eq-dk}). A comparison between $\left( 1 - 3k_B/(2c_v) \right)$ using Eq.~(\ref{eq-cvfit}) and MD is shown in Fig.~\ref{fig-cvs}. It is clear that Eq.~(\ref{eq-cvfit}) works well for approximating this quantity. Therefore, the two limiting expressions  for bulk viscosity, Eq.~(\ref{eq-weak-model}) and Eq.~(\ref{eq-strong-model}), can now be solved. 

Good agreement in the appropriate limits is shown in Fig.~\ref{fig-model}. Of course, the analytic formula for the asymptotic limits of $\Gamma$ fail in the intermediate region around $\Gamma \sim 1$. To bridge the two limits, we use a logarithmic sigmoid function, which provides the desired behavior of growing smoothly from 0 to 1 as $10^{-2} < \Gamma < 10^2$. It takes the form  
\begin{equation}
    \sigma(\Gamma) = \frac{1}{1+e^{-c_1\log(\Gamma+c_2)}}.
    \label{eq-sigma}
\end{equation}
where $c_1 = 12.5$ and $c_2 = 0.5$ are fit parameters. 
This can be used to create the following expression for bulk viscosity 
\begin{equation}
    \eta_v = \eta_{v}^{\textrm{wc}} [1 - \sigma(\Gamma)] +  \eta_{v}^{\textrm{sc}} \sigma(\Gamma), \label{eq-model}
\end{equation}
which when cast in dimensionless form ($\eta_v^* = \eta_v/mna^2\omega_p$) is
\begin{equation}
    \eta_v^* = \frac{1}{18\Gamma} \left( 1 - \frac{3}{2\tilde{c_v}}\right) \left[ 0.38(1 - \sigma(\Gamma)) + 0.88 \sqrt{\frac{\pi}{2}} \sigma(\Gamma)\right].
    \label{eq-final}
\end{equation}
Here, $\tilde{c}_v$ is evaluated using Eq.~(\ref{eq-cvfit}) and $\sigma(\Gamma)$ using Eq.~(\ref{eq-sigma}).

Equation~(\ref{eq-final}) provides an algebraic expression for bulk viscosity in the OCP. Results of this expression are shown in Fig.~\ref{fig-model}. Good agreement is observed across all coupling strengths, so Eq.~(\ref{eq-final}) can be used to quickly evaluate bulk viscosity in the OCP. 

\subsection{Yukawa One-Component Plasma \label{sec-yocp}}
\begin{figure}
    \centering
    \includegraphics[width=8.5cm]{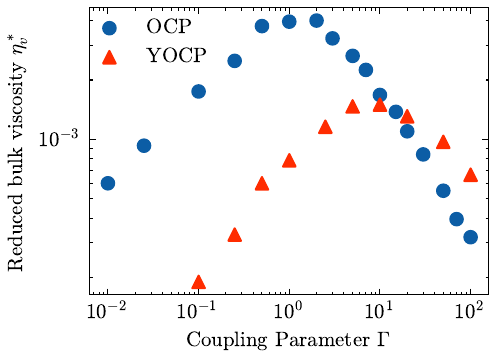}
    \caption{MD data of bulk viscosity of the OCP and YOCP with $\kappa = 1$.}
    \label{fig-yocp}
\end{figure}

In some strongly coupled plasmas, electrons screen the ion-ion interaction~\cite{dusty_review, Killian_2007}. In such a systems, the ion-ion interaction is best modelled using a screened Coulomb, i.e., Yukawa potential, of the form 
\begin{equation}
    \phi_Y(r) = \frac{q^2}{4 \pi \epsilon_0 r} e^{-r/\lambda_{\mathrm{sc}}},
\end{equation}
where $\lambda_{\mathrm{sc}}$ is the screening length.
A system of particles interacting through the Yukawa potential is known as the Yukawa one-component plasma (YOCP) model~\cite{hamaguchi_1994}. The YOCP is characterized by two dimensionless parameters, the Coulomb coupling parameter $\Gamma$ and the screening parameter 
\begin{equation}
    \kappa = \frac{a}{\lambda_{\mathrm{sc}}}.
\end{equation}
The focus of this study is the OCP. However, it is worth understanding the impact of electron screening on bulk viscosity. For this reason, MD simulations were run of the YOCP with a screening parameter of $\kappa = 1$. Data is shown in Fig.~\ref{fig-yocp} and provided in Tab.~\ref{tab-yocp}.

\begin{table}[]
    \centering
    \begin{tabular}{c|c||c|c}
        $\Gamma$ & $\eta_v^*$ & $\Gamma$ & $\eta_v^*$  \\ \hline
        0.1 &  $1.9 \times 10^{-4}$ & 5 & $1.5 \times 10^{-3}$ \\
        0.25 & $3.3 \times 10^{-4}$ & 10 & $1.5 \times 10^{-3}$ \\
        0.5 & $6.0 \times 10^{-4}$ & 20 & $1.3 \times 10^{-3}$ \\
        1 & $7.8 \times 10^{-4}$ & 50 & $9.7 \times 10^{-4}$ \\
        2.5 & $1.2 \times 10^{-3}$ & 100 & $6.6 \times 10^{-4}$ \\
    \end{tabular}
    \caption{Reduced coefficient of bulk viscosity in the YOCP at $\kappa = 1$ computed using equilibrium molecular dynamics under the Green-Kubo formalism.}
    \label{tab-yocp}
\end{table}

It is observed that the bulk viscosity peaks at a larger $\Gamma$ value in the YOCP than in the OCP. This occurs because electron screening weakens the ion-ion interaction, such that the dynamics and structure of the YOCP at $\Gamma \sim 10$ and $\kappa = 1$ closely resembles that of $\Gamma \sim 1$ in the OCP. It can also be seen that the bulk viscosity is smaller in the YOCP. This happens because the excess heat capacity of ions is smaller due to electron screening. This type of behavior is common when comparing the OCP and YOCP. For example, one can see an identical shift when looking at the shear viscosity~\cite{Murillo_2000,Daligault_2014}. 

\subsection{Frequency-dependent bulk viscosity \label{sec-freq}}

\begin{figure}
    \centering
    \includegraphics[width=8cm]{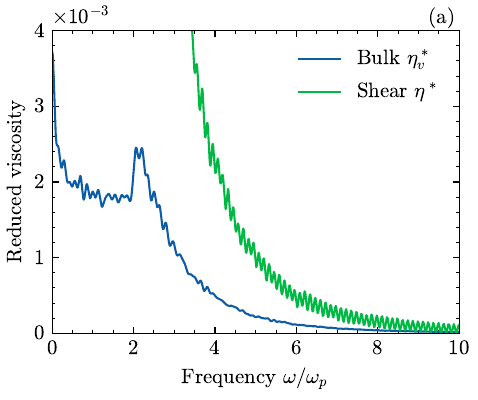}
    \includegraphics[width=8cm]{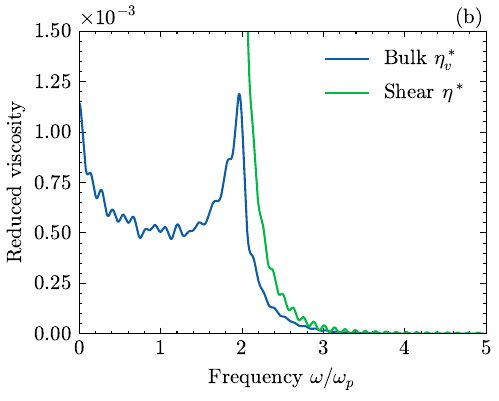}
    \includegraphics[width=8cm]{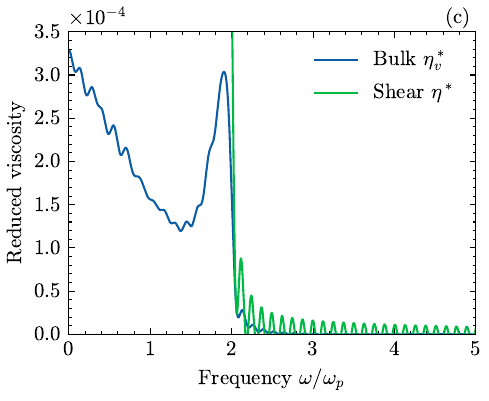}
    \caption{Reduced bulk viscosity and shear viscosity coefficients as a function of frequency for (a) $\Gamma = 1$, (b) $\Gamma = 20$, and (c) $\Gamma = 100$.}
    \label{fig-freq}
\end{figure}

Transport coefficients are typically computed in the hydrodynamic limit ($\omega \rightarrow 0$ and $k \rightarrow 0$), meaning over sufficiently long time and spatial scales to correspond to local thermodynamic equilibrium. However, hydrodynamics can be generalized to shorter space and timescales where transport coefficients are frequency dependent~\cite{Hansen_McDonald,Donko_2010}. 
For example, if one were to probe a fluid at a timescale faster than the relaxation time associated with bulk viscosity, bulk viscous dissipation would still occur, but to a lesser extent. This is particularly significant in molecular fluids, where bulk viscosity can have several associated relaxation times, each of which occur on different timescales~\cite{Sharma_review}. 
Since the relaxation times associated with bulk and shear viscosity can differ by several orders of magnitude, particularly at weak coupling, it is interesting to investigate if a timescale exists where bulk viscosity exceeds shear viscosity.

The frequency-dependent bulk viscosity and shear viscosities can be evaluated using the associated  Green-Kubo relations~\cite{Swanzig_1965}
\begin{subequations}
\begin{align}
    \eta_v(\omega) &= \frac{\mathcal{V}}{k_\textrm{B} T} \int_0^\infty dt  \left< \delta p(0) \delta p(t)\right> e^{i \omega t}     \label{eq-freq-bulk}  \\
    \eta(\omega) &= \frac{\mathcal{V}}{6 k_\textrm{B} T} \sum_{i=1}^3 \sum_{\stackrel{j=1}{j\neq i}}^3 \int_0^\infty dt  \left< \Pi_{ij}(0) \Pi_{ij}(t)\right> e^{i \omega t}.   \label{eq-freq-shear}
\end{align}
\end{subequations}
The imaginary part of of $\eta_v(\omega)$ and $\eta(\omega)$ is associated with solid-like (rigid and elastic) behavior that occurs on fast timescales~\cite{Hansen_McDonald}. The real part represents viscous dissipation and will be our focus here.

MD simulations proceeded in the same fashion as described in Sec.~\ref{sec-md}. Results at three separate $\Gamma$ values are shown in Fig.~\ref{fig-freq}. In all cases, the shear viscosity has a much larger value in the low frequency limit, consistent with the values corresponding to the hydrodynamic limit from Fig.~\ref{fig-model}. The relative magnitudes of bulk and shear viscosity become comparable on frequencies exceeding approximately $2\omega_p$, but the bulk viscosity remains smaller. 
The values only become approximately equal when both coefficients are consistent with zero at the accuracy of the simulations. Therefore, regardless of the timescale at which one observes, viscous dissipation due to shear viscosity appears to exceed that of bulk viscosity in the OCP.

\section{Conclusion \label{sec-conclusion}}

In this work, bulk viscosity of the one-component plasma was computed using MD simulations. It was shown that bulk viscosity peaks when there is a balance of kinetic and internal energy, corresponding to a coupling parameter $\Gamma$ of around 1. An understanding of the physics in the weakly coupled and strongly coupled limits was presented and used to construct a model that spans the entire range of values. Good agreement was obtained across all coupling strengths. The impact of electron screening was also studied using MD simulations of the YOCP. It was shown that screening reduces the bulk viscosity and shifts the peak to a larger $\Gamma$ value. Finally, the frequency-dependence of bulk viscosity and shear viscosity was computed. It was shown that even over very fast timescales, shear viscosity is larger than bulk viscosity in the OCP. 
Therefore, we can conclude that processes like sound wave dissipation that depend on the longitudinal viscosity, $b = \frac{4}{3}\eta + \eta_v$, are unlikely to be significantly influenced by bulk viscosity in the OCP. 

\begin{acknowledgements}

The authors thank Lucas Babati for providing the HNC code and for helpful conversations on this study. The authors also thank Dr. Louis Jose for helpful conversations on the study. This research was supported by the US Department of Energy under Award No. DE-SC0022201, and by computational resources and services provided by Advanced Research Computing (ARC), a division of Information and Technology Services (ITS) at the University of Michigan, Ann Arbor.
\end{acknowledgements}

\bibliography{apssamp}

\end{document}